\documentclass[final,5p,times,twocolumn,number]{elsarticle}

\usepackage[utf8]{inputenc}
\usepackage[T1]{fontenc}
\usepackage{amssymb}
\usepackage{lipsum}
\usepackage{graphicx}
\usepackage{natbib}
\usepackage{bm}
\usepackage[dvipsnames]{xcolor}
\usepackage{hyperref}
\usepackage{lineno}
\usepackage{amsmath}

\journal{Physics Letters B}

\begin{document}

\begin{frontmatter}



\title{Constraints on the Inert Doublet Model of dark matter with very high-energy gamma-ray observatories}


\author[a,b]{Lucca Radicce Justino}
\ead{luccarj2024@cau.ac.kr}
\affiliation[a]{
  organization={Instituto de Física de São Carlos, Universidade de São Paulo (IFSC--USP)},
  addressline={av. Trabalhador São-carlense, 400},
  city={São Carlos},
  postcode={13566-590},
  state={SP},
  country={Brasil}
}

\affiliation[b]{
  organization={Department of Physics, Chung-Ang University},
  city={Seoul},
  postcode={06974},
  country={Korea}
}

\author[c]{Clarissa Siqueira}
\ead{csiqueira@on.br}
\affiliation[c]{
  organization={Observatório Nacional},
  city={Rio de Janeiro},
  postcode={20921-400},
  state={RJ},
  country={Brasil}
}

\author[a]{Aion Viana}
\ead{aion.viana@ifsc.usp.br}

\begin{abstract}
We investigate constraints on the Inert Doublet Model (IDM) that features a scalar dark matter candidate, using data from recent and future gamma-ray observatories. The relevance of the model for indirect searches of dark matter stems from two key features: first, in the high mass regime, IDM can achieve the correct dark matter relic abundance for masses between approximately 500 GeV and 25 TeV, aligning perfectly with the energy sensitivity of Imaging Atmospheric Cherenkov Telescopes. Second, this regime is dominated by co-annihilation processes, which elevate the thermal relic velocity-weighted annihilation cross section to the range of $0.5$ to $1.0 \times 10^{-25}$ cm$^3$ s$^{-1}$, thereby enhancing the potential gamma-ray signal from dark matter annihilation. Analyzing the recent H.E.S.S. Inner Galaxy Survey, we find that dark matter particle masses within the 1-8 TeV range are excluded by current data, assuming a benchmark cuspy Einasto profile. Furthermore, we project that the Cherenkov Telescope Array Observatory (CTAO) will comprehensively probe the remaining viable parameter space of the IDM under the same assumptions. Our findings are further examined in light of recent theoretical constraints, collider searches, and direct detection results from the LUX-ZEPLIN experiment.
\end{abstract}



\begin{keyword}
IDM model \sep indirect detection \sep Gamma-ray searches 



\end{keyword}

\end{frontmatter}




\section{Introduction}

The true nature of dark matter (DM) remains one of the most pressing questions in modern physics, driving extensive experimental efforts across multiple platforms. Colliders, direct-detection experiments, and indirect-detection methods have been used to search for signals of DM particles across a wide range of masses. Among the most compelling candidates are weakly interacting massive particles (WIMPs), which are theorized to have masses in the GeV to TeV range~\cite{Jungman:1995df,Garrett:2010hd}. Despite numerous attempts, no definitive detection has yet been achieved, leading to stringent constraints on various extensions of the Standard Model (SM) that predict WIMP candidates within this mass range~\cite{Arcadi:2017kky,Arcadi:2024ukq}. As a result, the exploration of multi-TeV DM scenarios has become increasingly important. However, probing this mass range presents significant challenges for direct detection experiments, which lose sensitivity at such high masses, and for colliders, which struggle to reach the required energy scales. Consequently, alternative methods, such as indirect detection via very-high-energy (VHE) gamma rays, have become crucial in the search for WIMPs~\cite{CTAConsortium:2017dvg}.




In indirect detection, it is hypothesized that DM may self-annihilate in dense astrophysical environments, producing SM particles that could be observed on Earth. These include charged cosmic rays, neutrinos, and gamma rays. Charged particles, such as electrons, positrons, and protons, are deflected by astrophysical magnetic fields, making it difficult to trace them back to their origin. Neutrinos, although capable of traveling vast distances unimpeded, are notoriously challenging to detect due to their weak interactions with matter. In contrast, gamma rays offer distinct advantages: they are not deflected by magnetic fields, furthermore, they are only mildly affected by absorption within the Galaxy, making them valuable messengers for indirect searches for DM. With current technology, gamma rays can be detected with high precision, providing a crucial window into possible DM interactions. Several instruments, including Fermi-LAT~\cite{Fermi-LAT:2025gei}, H.E.S.S.~\cite{HESS:2022ygk}, MAGIC \cite{MAGIC:2021mog}, VERITAS \cite{VERITAS:2017tif}, HAWC \cite{HAWC:2023owv}, and LHAASO \cite{LHAASO:2022yxw}, have been searching for a VHE gamma-ray signal from generic WIMPs.  
Notably, H.E.S.S. observations of the Inner Galaxy halo have provided the most stringent constraints on multi-TeV DM, excluding annihilation cross sections below the thermal relic value \cite{aghanim2020planck} for DM masses between 200 GeV and a few TeV, particularly for annihilation via the $\tau^+\tau^-$ channel \cite{HESS:2022ygk}. However, it is important to note that these limits were obtained using a model-independent approach, where the annihilation is assumed to happen with a 100\% branching ratio into SM particle/antiparticle pairs. 
The next generation of gamma-ray observatories, such as the Cherenkov Telescope Array Observatory (CTAO) \cite{CTAConsortium:2017dvg} and the Southern Wide-field Gamma-ray Observatory (SWGO) \cite{swgo}, are expected to enhance sensitivity to DM signals by an order of magnitude, especially for DM masses above the TeV scale. 

In this work, we derive the most recent indirect detection limits for one of the simplest yet specific DM models, the Inert Doublet Model (IDM). The IDM is a minimal extension of the Standard Model (SM), where a new Higgs doublet is added. This new doublet is stabilized by a discrete $Z_2$ symmetry, which allows the lightest neutral particle to be a DM particle candidate. In the IDM, the observed abundance of DM can be obtained in two distinct regimes: the low-mass regime ($m_{DM} < m_W, m_Z$), and the high-mass regime ($m_{DM} > 500 \, \text{GeV}$) \cite{belyaev2018anatomy}. The former is already highly constrained by collider and direct detection data. Therefore, it is not the current focus of indirect searches \cite{Ilnicka:2015jba,Kalinowski:2018ylg,belyaev2018anatomy,Eiteneuer:2017hoh,Heisig:2018kfq,Tsai:2019eqi,Ghosh:2021noq,Ghosh:2024boo}. On the other hand, the high mass regime is very promising for indirect detection searches for two important reasons. First, in this regime, the DM particle is expected to annihilate, producing gamma rays in the TeV energy scale. Second, the annihilation cross section is enhanced by co-annihilation effects in the high mass limit, leading to a higher gamma-ray signal. These characteristics make the IDM one of the best DM models to probe in indirect searches with VHE gamma-ray telescopes such as H.E.S.S. and CTAO.

In previous studies \cite{Queiroz:2015utg, garcia2016probing, duangchan2022cta}, indirect dark matter limits were typically derived assuming pure annihilation channels, using simplified spectral shapes to estimate sensitivities. In contrast, in this work, we compute the full continuum spectra predicted for each model scenario, based on detailed simulations of all relevant annihilation channels and their branching fractions. These spectra are then directly injected into a binned two-dimensional likelihood analysis (energy and spatial dimensions) to derive robust constraints on the annihilation cross section.

The present work advances beyond previous studies in several key aspects:
\begin{itemize}
\item We derive new indirect-detection limits on the IDM using the most recent H.E.S.S. data, with a significant impact on the allowed parameter space;
\item We incorporate the current strongest direct-detection limits from LUX-ZEPLIN (LZ);
\item We compute projected sensitivities for CTAO using the complete simulated spectra of the dominant annihilation channels, rather than simplified pure-channel assumptions;
\item We perform a binned 2D (in energy and space) likelihood analysis for CTAO sensitivities, using the latest publicly available IRFs.
\end{itemize}

Finally, we combine these three constraints to highlight their complementarity and identify the regions of parameter space that remain viable for the model.


\section{The Inert Doublet Model}

The IDM is a minimal extension of the scalar sector of the SM where a new Higgs doublet $H_2 = \left( H^+ , \left(H + i \, A\right) /\sqrt{2}     \right)$ that does not participate in the electroweak symmetry breaking is added \cite{belyaev2018anatomy}. This new doublet is odd under a discrete $Z_2$ symmetry, while all the SM particles are even. This symmetry stabilizes the lightest particle, allowing one of the neutral components to be the candidate for the DM particle. This new doublet has no Yukawa couplings to the fermions, preventing tree-level neutral flavour-changing interactions.   

Beyond the kinetic term, determined by the $SU_L (2) \times U (1)_Y$ electroweak symmetry, the new doublet interacts with the original Higgs doublet ($H_1$) through the potential term given by
%
%
\begin{equation}
\begin{split}
    V(H_1, H_2) = \mu_{1}^2|H_1|^2 + \mu_{2}^2 |H_2|^2 + \lambda_1 |H_1|^4 + \lambda_2 |H_2|^4 \\ + \lambda_3 |H_1|^2 |H_2|^2 + \lambda_4 |H_1^{\dagger} H_2|^2 + \lambda_5 Re \left[ \left(H_1^{\dagger} H_2\right)^2 \right],
\end{split}
\label{eq.IDM_potential}
\end{equation}
%
%
where $\mu_2^2 > 0$ and the parameters $\lambda_i$'s are real. This term also determines the masses of the scalar sector after the spontaneous symmetry breaking. The masses of the Higgs boson $h$, the charged scalar $H^+$, the neutral scalar $H$ and the neutral pseudo-scalar $A$ bosons will be
\begin{equation}
\begin{split}
    m_{h}^2 = 2\lambda_1  v^2, \, m_{H}^2 = \mu_{2}^2 + \lambda_{345} v^2, \\
    m_{A}^2 = \mu_{2}^2 + \bar{\lambda}_{345} v^2 , m_{H^{\pm}}^2 = \mu_{2}^2 + \frac{1}{2} \lambda_3 v^2,
\end{split}
\label{eq.IDM_masses}
\end{equation}
where $\lambda_1$ and $\mu_1^2$ are SM parameters and $v = - \mu_1^2/ \lambda_1 = 246 \, \text{GeV}$ is the vacuum expectation value (VEV) of the Higgs field. The masses of the inert doublet particles depend on the two physical couplings called $\lambda_{345}$ and $\bar{\lambda}_{345}$, 
defined as
\begin{equation}
    \lambda_{345} = \frac{1}{2}\left(\lambda_3 + \lambda_4 + \lambda_5\right), \quad \bar{\lambda}_{345} = \frac{1}{2}\left(\lambda_3 + \lambda_4 - \lambda_5\right).
\end{equation}
Eqs.~\eqref{eq.IDM_masses} determine the hierarchy of masses of the IDM. Either $H$ or $A$ can be the lightest particle and the DM candidate, since they yield the same phenomenology. Here, we assume that $H$ is the lightest particle, without loss of generality. 


As shown in \cite{belyaev2018anatomy}, there are two regimes where the observed relic abundance $ \Omega \, h^2 = 0.1200 \pm 0.0012 $ \cite{aghanim2020planck} is obtained: the low-mass regime ($ m_H \lesssim m_W,m_Z  $) and the high mass regime ($ m_H \gtrsim 500 \, \text{GeV} $). The low-mass regime is already strongly constrained by collider and direct-detection data; see, for instance, \cite{tytgat2006inert,belyaev2018anatomy,Heisig:2018kfq,Eiteneuer:2017hoh}. The intermediate-mass range ($ m_W,m_Z < m_H < 500 \, \text{GeV} $) is excluded by relic abundance because it is very suppressed by annihilation into the electroweak gauge bosons \cite{belyaev2018anatomy} \cite{Queiroz:2015utg}. The high mass regime can only yield the right abundance if the mass splittings of the partner particles are very small ($\Delta_{o,+} = m_{A,H^\pm} - m_{H} \lesssim 10 \, \text{GeV} $). In this case, the thermal production of DM is dominated by co-annihilation effects \cite{profumo2019introduction}. For a model with many particles, the heavier partners can annihilate with each other or even with the DM particle (co-annihilation processes). For a high degeneracy of masses, their abundances will freeze out around the same time, and the heavier particles will decay to the DM particle. In this scenario, the final abundance can be enhanced, and it is ruled by the effective cross section, which is a Boltzmann-weighted average between all the (co-)annihilation cross sections~\cite{profumo2019introduction}. The co-annihilation regime establishes that the DM annihilation cross section must be higher than the typical WIMP value, providing the same DM abundance. This means that the gamma-ray flux from annihilation in the IDM is also higher than usually expected, making indirect detection a very promising method. Another effect that may enhance the detection signal is the so-called Sommerfeld Enhancement (SE). This is a non-perturbative effect that occurs when the velocities of the initial-state particles are non-relativistic, resulting in a resonant quasi-bound state. In the IDM, this occurs when the electroweak gauge bosons and the SM Higgs are light compared to the dark matter particle ($m_H \gtrsim 1\ \mathrm{TeV}$). The impact of the SE in the IDM has been investigated in \cite{garcia2016probing}. In our analysis, we adopt the conservative tree-level annihilation cross sections as the baseline for our limits. Still, we also compare our results with estimations of the SE's possible influence in the exclusion limits in Section~\ref{sec:results}.

In the IDM, the DM particle can be probed by the typical methods of WIMP detection. Collider and direct detection searches are particularly effective in the low-mass regime. For example, for the former, extensive analyses of the IDM, including electroweak precision tests (EWPT), Higgs decay data, mono-X and vector boson fusion plus missing energy searches, have been performed in Ref.~\cite{belyaev2018anatomy}. In the high mass regime, collider constraints become weak; thus, direct and indirect detections are more relevant. We will also apply the most recent direct-detection limits from the LUX-ZEPLIN experiment. The limits from indirect detection in the high-mass regime were deeply investigated in Refs.~\cite{Queiroz:2015utg,garcia2016probing,Cuoco:2017iax,duangchan2022cta}. The main relevant channels for DM annihilation in the IDM were found to be the diboson final states ($W^+ W^-$, $Z Z$, and $h h$). These works have shown a strong potential of the CTAO to probe the IDM in the Galactic Center region, excluding almost all scenarios. In Ref.~\cite{garcia2016probing}, other effects such as electroweak corrections were analyzed. The sensitivity of CTAO to the IDM for dwarf galaxies was projected in Ref.~\cite{duangchan2022cta}, where it was found that Draco and Sculptor do not provide strong enough limits to probe the IDM thermal-relic velocity-weighted annihilation cross section. The effect of astrophysical uncertainties on the small-mass regime of IDM was studied in Ref.~\cite{Benito:2016kyp}, where limits from direct detection, dwarf galaxies, and the GC excess were analyzed.  

\section{Methodology}

\subsection{Scanning the Inert Doublet Model}

To assess the viability of the IDM, we scanned the model's parameter space and checked the relevant theoretical and experimental constraints \cite{belyaev2018anatomy}. The IDM has 7 parameters, as seen in Eq.~\eqref{eq.IDM_potential}. Two of them, $\mu_1^2$ and $\lambda_1$, are determined by the Higgs boson mass $m_h$ and the Higgs VEV $v$. Regarding the DM phenomenology, the $\lambda_2$ parameter appears only in loop processes; thus, it plays no significant role in the DM observables, so we fixed a value for it. Therefore, there are four physical parameters of the IDM that must be probed. They can be taken as ``$m_H$'', ``$\Delta_+$'', ``$\Delta_o$'', ``$\lambda_{345}$'', i.e., the heavy scalar mass (our DM candidate), the mass splittings of the co-annihilating particles, and the physical coupling of the DM particle with the SM.

We performed two scans across the parameter space to obtain samples of possible IDM scenarios. A random sampling of these parameters generated the first scan to identify the points in the $\left\langle \sigma_{\text{ann}} \, v \right\rangle \times m_H$ space that satisfy the correct relic abundance, along with the theoretical, collider, and direct detection constraints. Evaluating the indirect detection limits across the IDM parameter space is computationally expensive. For that reason, a second scan was performed by selecting benchmarks to determine the limits of indirect detection with gamma rays. Both scans use the following parameter ranges:
\begin{equation}
\begin{array}{c}
    0 < \lambda_{345} < 2 \pi,  \\
    300 \, \text{GeV} < m_H < 30  \,\text{TeV}, \\
    0.5 \, \text{GeV} < \Delta_+ < 10 \, \text{GeV}, \\
    0.5 \, \text{GeV} < \Delta_o < 10 \, \text{GeV}. \\
\end{array}
\label{eq.scan_IDM_range}
\end{equation}

In these ranges, constraints such as vacuum stability, LEP electroweak precision data, and LHC Higgs decay are automatically preserved. It was checked that these ranges agree with EWPT parameters $S$ and $T$ within 1 and 2 standard deviations of coverage, respectively. Next, we closely follow the steps described in Ref.~\cite{belyaev2018anatomy}, where further theoretical and observational constraints are used as subsequent cuts in the parameter space. They are: 
\begin{itemize}
    \item Cut-1) constraints from unitarity \cite{belyaev2018anatomy},  and inertness, i.e., $\mu_2^2 = m_H^2 - \lambda_{345} v^2 > 0$;
    \item Cut-2) the measured relic abundance of DM by the Planck satellite \cite{aghanim2020planck} $\Omega \, h^2 = 0.1200 \pm 0.0012$;
    \item Cut-3) upper limits at 90 \% C.L. from the DM direct detection experiment LUX-ZEPLIN (LZ) \cite{aalbers2023first}.
\end{itemize}

After the scan is performed and cut-1 is applied, we use the micrOMEGAs 5.3.35 software \cite{Belanger:2014vza,Belanger:2018ccd,Alguero:2022inz} to evaluate the relic abundance for each scenario allowed by the scan, thereby determining cut-2. MicrOMEGAs is a software that computes the relic density for a stable massive particle by evaluating the relevant annihilation and co-annihilation cross sections and solving the Boltzmann equation numerically. Finally, cut-3 is applied using the expression of the WIMP-nucleus cross section (see Ref.~\cite{barbieri2006improved}). 

\subsection{Indirect DM detection using gamma rays}


The gamma-ray flux from the annihilation of WIMP particles in a certain DM distribution is given by
\begin{equation}
    \frac{d\Phi}{dE} = \frac{\left< \sigma_{\text{ann}} v \right>}{8 \pi m_{DM}^2} \sum_i B_i \frac{dN_i}{dE} J \left( \Delta \Omega \right),
\end{equation}
where the flux depends on the particle physics parameters: the DM particle mass $m_{DM}$, the annihilation velocity-weighted cross section $\left\langle \sigma_{\text{ann}} v \right\rangle$, the spectrum per annihilation $\frac{dN_i}{dE}$ of the primary channel $i$, and the branching ratio of each primary channel $B_i$. The flux also depends on the $J$-factor, which encodes the astrophysical DM distribution $\rho_{DM}$ of the target. For a given target, the $J$-factor is defined as
\begin{equation}
    J \left( \Delta \Omega \right) = \int_{\Delta \Omega} d \Omega \int_{\text{l.o.s.}} d s \rho_{DM}^2 \left( r \left( s, \Omega \right) \right),
    \label{eq.J-factor}
\end{equation}
where $r$ is the distance from the center of the DM halo, $s$ is the distance from the observer, and $\Omega$ denotes the direction in the sky for a given region with solid angle $\Delta \Omega$. This factor depends on the object's astrophysical type and the profile model for $\rho_{DM}(r)$. In this paper, we have chosen a cuspy Einasto profile for the Milky Way halo with parameters taken from Refs.~\cite{abdallah2016search, Abramowski:2011hc}, to consistently compare with previous works conducted using H.E.S.S.~\cite{abdallah2016search} data and the predictions for CTAO~\cite{CTAConsortium:2017dvg,CTA:2020qlo}. The profile considered here has parameters $\rho_s$ = 0.079, $r_s$ = 20~kpc and $\alpha = 0.17$. The parameter $\rho_s$ was calculated by assuming a local DM density of 0.39 GeV/cm$^3$ \cite{Catena:2009mf,Benito:2019ngh,2019JCAP...09..046K}. It is important to note, however, that the DM density distribution of the Milky Way is not fully known and is subject to sizable uncertainties, especially in its innermost regions. In Refs.~\cite{Benito:2016kyp}, \cite{Benito:2019ngh}, and \cite{Benito:2020lgu}, it has been shown that a factor of almost 10 of uncertainty on the J-factor is expected, which would in turn affect the gamma-ray flux and limits evaluated here by roughly the same factor.

The gamma-ray flux resulting from DM annihilation can be detected by gamma-ray observatories operating in the GeV-TeV energy range. When observed, this flux is convolved with the telescope's instrument response functions (IRFs). The signal can be identified as an excess over the astrophysical background using a likelihood-ratio test. The background can be estimated either through modeling or by directly observing a specific region of the sky. The former approach yields robust results but introduces significant systematic errors. The latter, known as the ON/OFF method, involves selecting a region of interest (ROI), called the ON region, and a control region, called the OFF region, from which the background is estimated. Sensitivity can be further enhanced through a two-dimensional binned analysis that considers both spectral and spatial information. The expected gamma events count $S$ associated with the DM annihilation is given by
\begin{equation}
\begin{split}
    {S}_{ij} \left( \left< \sigma_{\text{ann}} \, v \right> \right) = & \left< \sigma_{\text{ann}} \, v \right> \frac{T_{\text{obs}} J \left( \Delta \Omega_j \right)}{8 \pi m_{DM}^2} \times \\
    & \int_{\Delta E'_i} \int_0^\infty \frac{d N}{dE} A_{\text{eff}} (E) R(E,E') dE dE',
\end{split}
\label{eq.S_sigmav}
\end{equation}
where $i$ ($j$) is the index of the energetic (spatial) binning, $T_{\text{obs}}$ is the observation time, $A_{eff}$ is the effective area of the telescope, $R(E,E')$ represents the energy resolution, as the probability density function of observing an event at reconstructed energy $E'$ for a given true energy $E$, and $dN/dE$ is the total annihilation spectrum summed over all primary channels. In our work, the total spectrum for each scenario was calculated by the PPPC4DMID accessed through the micrOMEGAs code, and the cross section was calculated using CalcHEP, both in accordance with the tree-level formulas.  \cite{Belyaev:2012qa,Belanger:2018ccd,Alguero:2022inz}


When no signal is detected, upper limits can be established over the annihilation cross section \cite{Cowan:2010js}. The expected exclusion limits are determined by the log-likelihood ratio given by
%
\begin{equation}
    TS \left( \left< \sigma_{\text{ann}} \, v \right> \right) = - 2 \ln{\left(\frac{\mathcal{L} \left( \left< \sigma_{\text{ann}} \, v \right>  \right) }{\mathcal{L}\left( \left< \sigma_{\text{ann}} \, v \right> = 0 \right)}\right)},
    \label{eq.ts}
\end{equation}
where $\mathcal{L} \left( \left< \sigma_{\text{ann}} \, v \right>  \right) = \Pi_{ij} \mathcal{L}_{ij}$ is the joint likelihood of all the regions of interest (ROI) and energy bins. The likelihood for each bin is determined by the Poisson distribution of the ON and OFF regions, expressed as   
\begin{equation}
    \mathcal{L}_{ij} = \frac{\left( S_{ij} + B_{ij} \right)^{N_{ON ij}}}{N_{ON ij}!} e^{-\left( S_{ij} + B_{ij} \right)} 
    \frac{\left(  B_{ij} /\alpha \right)^{N_{OFF ij}}}{N_{OFF ij}!} e^{-\left( B_{ij}/ \alpha \right)},
\end{equation}
where $B_{ij}$ is the expected background and $N_{ON(OFF)ij}$ is the observed counts for each region. The parameter $\alpha$ is the ratio between the solid angle of the ON and OFF regions.

The test statistic of Eq.~\eqref{eq.ts} thus determines exclusion limits according to the chosen confidence level. Usually, an exclusion at 95 \% C.L.\footnote{This choice is different from the 90 \% C.L. for DD experiments but is standard for ID probes. Although these limits are not, strictly speaking, statistically comparable, we choose to rely on the DD results rather than recast the limits.} is considered, which implies a test statistic of $TS = 2.71$. The upper limit for the annihilation cross section of a DM particle of fixed mass $m_{DM}$ and annihilation spectrum $dN/dE$ is thus determined by 
\begin{equation}
    \left< \sigma_{\text{ann}} \, v \right>^{95\% CL} = TS^{-1} \left( 2.71 \right).
\label{eq.sigmav_95}
\end{equation}
Other effects can be taken into account, such as systematic error and a signal in the OFF regions, but here only statistical error and signal in the ON regions will be considered. The effect of systematic uncertainties has been studied in the context of H.E.S.S. observations of the Inner Galaxy region, and an impact of roughly $\sim 20$\% on the DM sensitivities has been reported~\cite{HESS:2022ygk}.  As for the impact of the contamination of our signal and control regions by the Galactic Diffuse Emission (GDE), it has been shown that in an ON-OFF analysis, if properly modeled, the GDE would degrade the CTAO sensitivity by a factor of $\sim$2~\cite{lefranc2015prospects}. 


A circular 1$^{\circ}$ region centered at the Galactic Center (GC) was considered for the analysis. Following the H.E.S.S. strategy, this area was further subdivided into seven concentric rings as our spatial ROIs, each 0.1$^{\circ}$ wide, starting at 0.3$^{\circ}$ from the GC. A mask within the galactic latitudes $b = \pm 0.3^{\circ}$ is applied to exclude gamma-ray emissions from the Galactic Plane at the TeV scale \cite{aharonian2004very,aharonian2006discovery,hess2016acceleration,viana2019searching}. The ON regions are chosen to observe the DM signals with the largest J-factor. The OFF regions, in turn, are chosen as reflected regions of the sky that are symmetric to the ON regions relative to the telescope pointing position~\cite{abdallah2016search}. This ensures that the IRFs will be the same at the same offset and observation time. 

\begin{figure*}[!ht]
    \centering
    \includegraphics[width=0.7\textwidth]{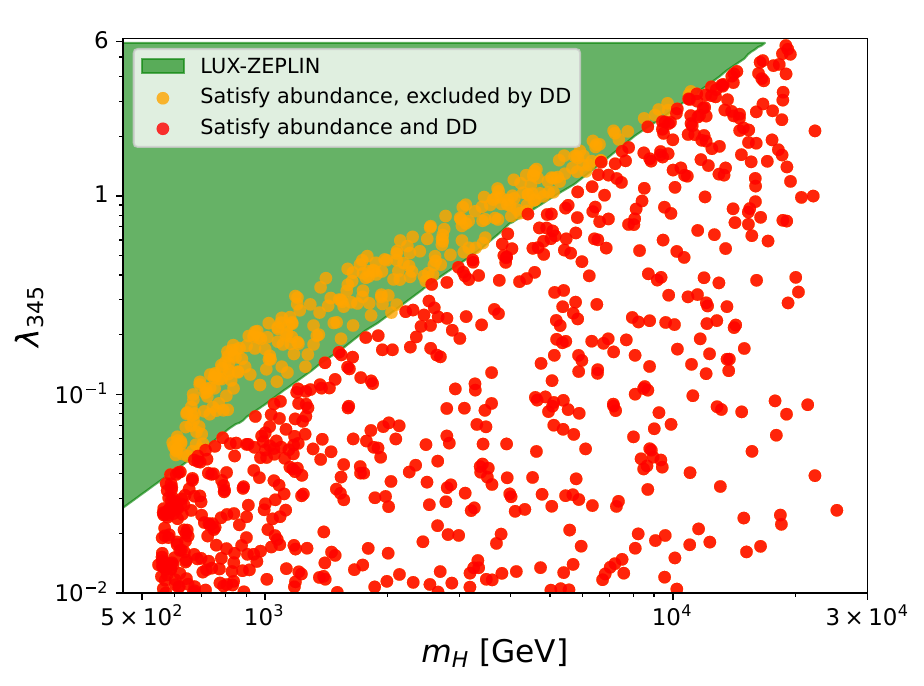}
    \caption{Random scan in the $\lambda_{345} \times m_H$ parameter space of the IDM. Each point represents a scenario that agrees with unitarity (cut-1) and relic abundance $\Omega \, h^2 =  0.1200$ (cut-2). Direct detection (DD) exclusion from LZ (cut-3) is represented in green. Yellow points satisfy relic abundance but are excluded by direct detection. Red points satisfy relic abundance and direct detection.}
    \label{fig.IDM_random_scan_exact}
\end{figure*}
\begin{figure*}[!ht]
    \includegraphics[width=1.1\textwidth]{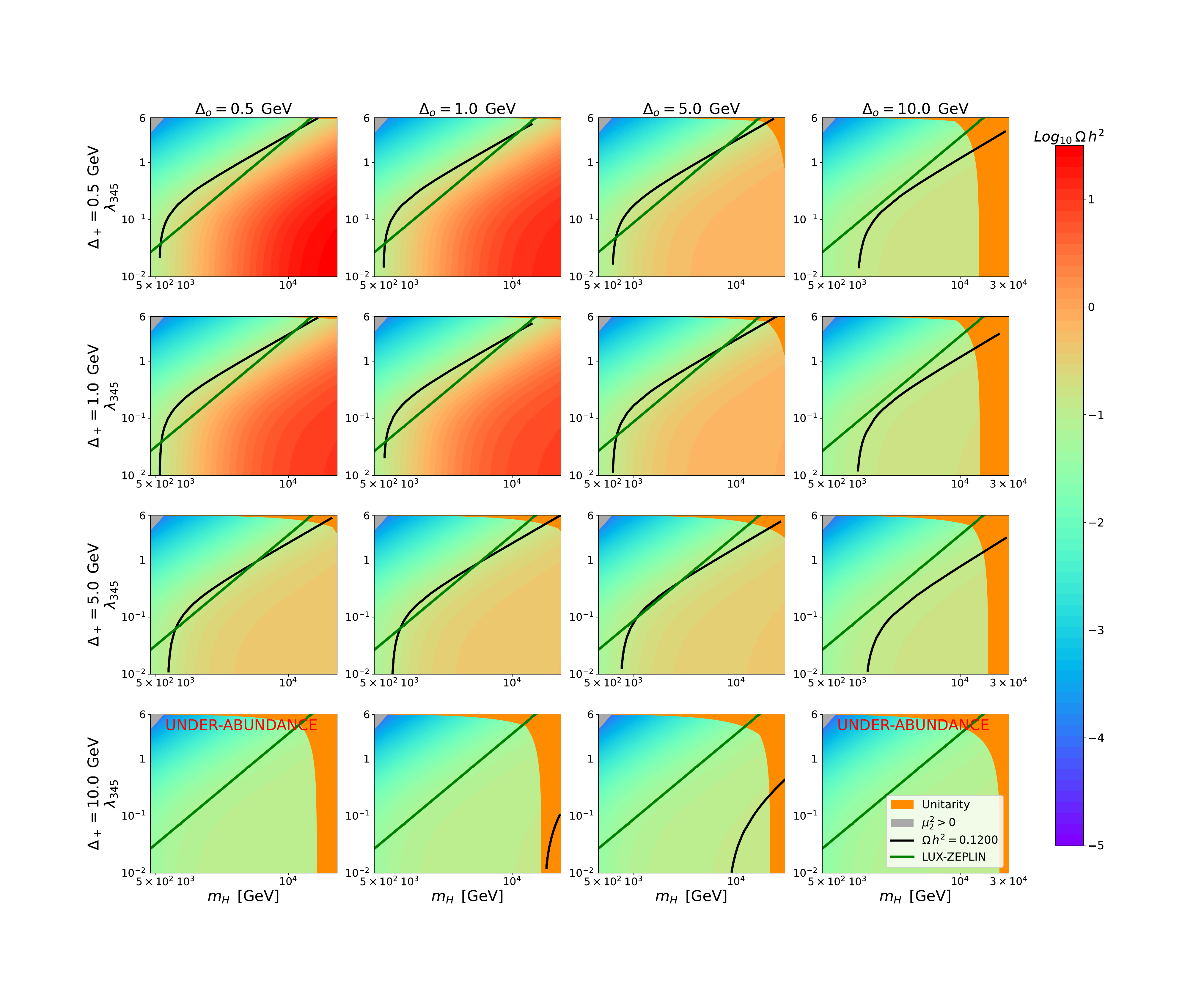}
    \caption{Color map of the relic abundance $\Omega \, h^2$ at logarithmic scale in the $\lambda_{345} \times m_H$ parameter space for the discrete scan. Each row corresponds to one of the mass splittings $\Delta_+ = \left( 0.5, 1, 5, 10 \right)  \text{GeV}$ and each column to one $\Delta_o$ in the same selection of values. Black lines: correct relic abundance. Green lines: LZ exclusion limits. Orange band: unitarity constraint. Grey band: inertness constraint. See the text for details.}
    \label{fig.IDM_discrete_scan}
\end{figure*}
In this work, since we do not have access to H.E.S.S. private analysis tools and data, the detected background event's distribution for each ROI and the instrument's effective area are extracted from Ref.~\cite{lefranc:tel-01374541}, which are the same as those used for the 10-year H.E.S.S. DM continuum searches~\cite{abdallah2016search}. We assume 545 hours of observation, which is equivalent to the H.E.S.S. Inner Galaxy Survey \cite{HESS:2022ygk} observation time and thus determine the counts $N_{ON}$ and $N_{OFF}$.


We also update the projected expected limits of the CTAO. In this case, the dataset was simulated using the updated IRFs that are publicly available on the CTAO website \cite{ctaperformance}. We use the more recent \textit{Alpha Configuration} IRFs, which can be downloaded through the file called \textit{prod5 version v0.1}. In this configuration, the southern array consists of 14 Medium-Sized Telescopes (MSTs) and 37 Small-Sized Telescopes (SSTs). In this case, 525 hours of observation are considered. The count $N_{OFF}$ is simulated using the hadronic isotropic background model, as given by the CTAO IRFs. It should be explicitly noted that these projected limits represent an optimistic baseline scenario assuming statistical uncertainties only.

For both H.E.S.S. and CTAO, the limits assuming 100\% annihilation into primary channels were re-obtained and found to be consistent with the published official results \cite{acharyya2021sensitivity,abdallah2016search,HESS:2022ygk} (see ~\ref{app:lim} for details). All codes and datasets used are available for download~\footnote{\url{https://github.com/RadicceJustino/IDM-indirect-detection.git}}.


%

\begin{figure*}[!ht]
    \includegraphics[width=1.1\textwidth]{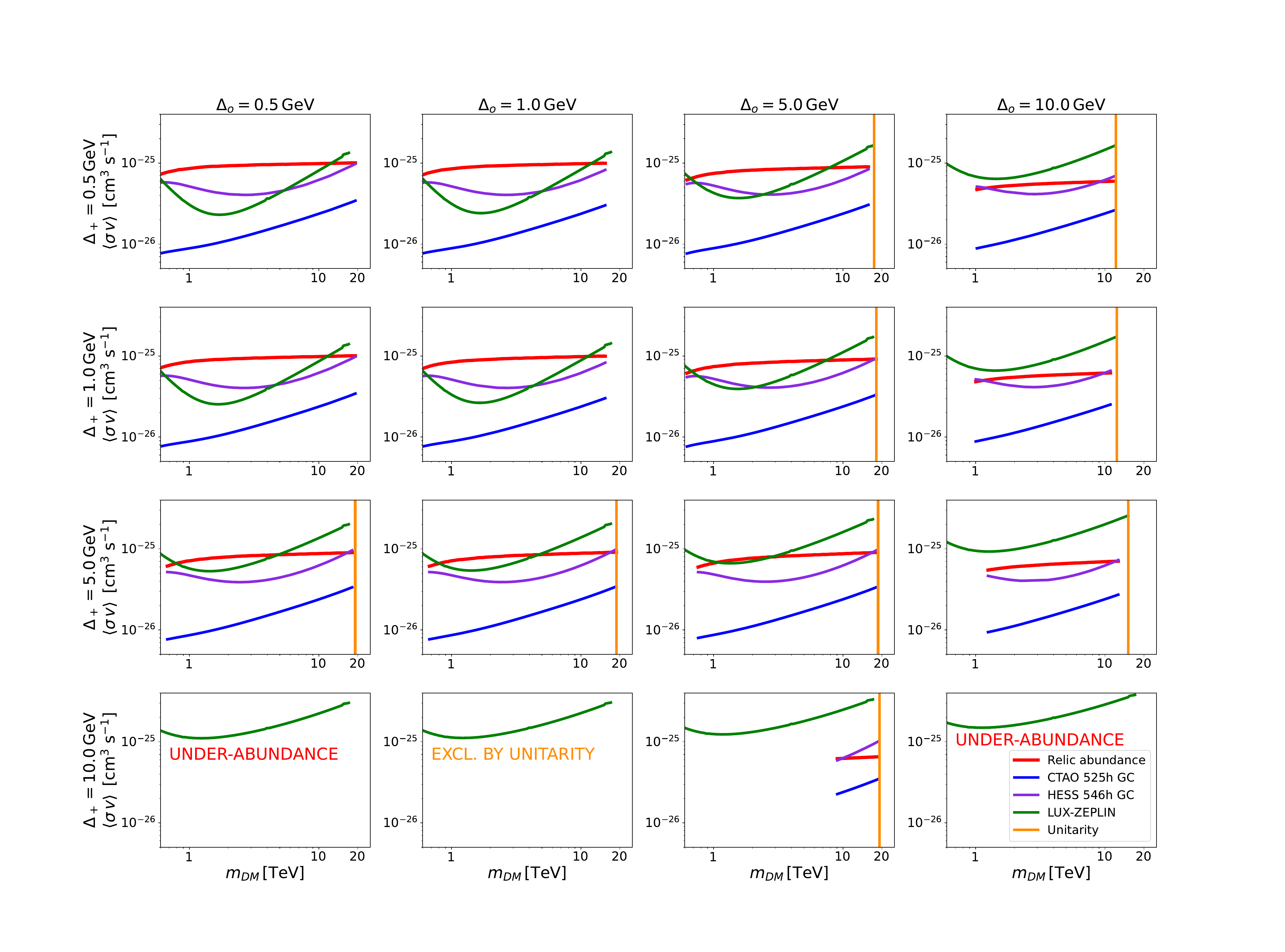}
    \caption{Compilation of constraints to the IDM in the annihilation velocity-weighted cross section versus DM particle mass ($\left< \sigma_{\text{ann}} \, v \right> \times m_{DM}$) space: relic abundance (red lines), direct detection exclusion limits at 90\% C.L. from LZ (green lines) and indirect detection expected limits at 95\% C.L. for H.E.S.S. (purple lines) and for CTAO (blue lines), both at the Galactic Center. Both the H.E.S.S. limits and the projected CTAO sensitivities are evaluated at the Galactic Center utilizing a benchmark cuspy Einasto profile. The CTAO curve represents an optimistic baseline projection using statistical errors only. Each row corresponds to one of the mass splittings $\Delta_+ = \left( 0.5, 1, 5, 10 \right)  \text{GeV}$ and each column to one $\Delta_o$ in the same selection of values. The unitarity upper bound for mass is shown in orange. The viable models are those on the red curves, which lie below the direct and indirect detection limits and to the left (below) the unitarity limits.} 
    \label{fig.IDM_limits_dicrete}
\end{figure*}

\section{Results and Discussion}
\label{sec:results}

\subsection{Scans on the parameter space of the Inert Doublet Model}

In the first scan, 250,000 scenarios were generated with parameters logarithmically sampled in the ranges displayed in Eq.~\eqref{eq.scan_IDM_range}. Then, cut-1 (unitarity) \cite{belyaev2018anatomy} was applied, and micrOMEGAs was used to evaluate the relic abundance. Cut-2 is then applied ($\Omega \, h^2 \approx 0.1200$) \cite{aghanim2020planck}. Cut-3 from LZ data provides, as upper limits, a curve $\lambda_{345} \sim m_H^{3/2}$ \cite{duangchan2022cta} which considerably restricts the available parameter space. The result is shown in Figure \ref{fig.IDM_random_scan_exact} where the axes are the scalar DM particle mass $m_H$ and the coupling $\lambda_{345}$. This picture shows how unitarity and relic abundance constrain the DM particle mass to be greater than $500 \, \text{GeV}$ and below $\approx 25 \, \text{TeV} $. Direct detection (DD) excludes some models with $m_H \lesssim 10 \, \text{TeV}$ (yellow points), leaving approximately 800 scenarios (red points) satisfying all constraints.  

For the evaluation of indirect detection limits, we follow the approach of Ref.~\cite{duangchan2022cta}, where some benchmarks are taken by fixing the mass splittings of the co-annihilating particles. We choose mass-splittings of $\Delta_o , \Delta_+ = 0.5,\, 1,\, 5$ and $10 \, \text{GeV}$, providing 16 benchmarks. For each benchmark, the mass $m_H$ was varied with logarithmic steps between $300 \, \text{GeV}$ and $30 \, \text{TeV}$, and the same was made for the coupling $\lambda_{345}$ between $10^{-2}$ and $2 \pi$. For each combination of the parameters, the abundance was evaluated by micrOMEGAs, and the points that provide the correct relic abundance ($\Omega \, h^2$) were selected, defining the \textit{correct abundance curves} $\lambda_{345} \left( m_H \right)$ for each benchmark. For each mass value $m_H$ of these curves, the annihilation gamma-ray spectra $ \frac{dN}{dE} = \sum_i B_i \frac{dN_i}{dE}$ and the branching ratios $B_i$ of the main channels were evaluated using micrOMEGAs. This final spectrum is then used as an input to evaluate the indirect detection limits for the benchmarks.  

The abundance maps for the discrete scan are shown in Figure \ref{fig.IDM_discrete_scan} together with unitarity and inertness constraints (cut-1). The correct abundance curves are represented in black (cut-2), while the direct detection LZ exclusion limits are in green (cut-3). From that, some properties of the IDM can be observed. Abundance increases to high DM masses $m_H$ or low couplings $\lambda_{345}$. The effect of co-annihilation is seen as a small mass splitting, implying an enhancement in abundance. It is also evident that direct detection and relic abundance are complementary for the IDM. This means that increasing direct-detection sensitivity in the future will further constrain the available scenarios. Unitarity constraints, shown in orange, limit the upper value for the mass $m_H$ up to $\approx 25 \, \text{TeV}$. The $\Delta_+ = 10 \, \text{GeV}$ row is excluded either by under-abundance or by unitarity, except for $\Delta_o = 5 \, \text{GeV}$.

\subsection{Indirect detection limits to the Inert Doublet Model}

Figure \ref{fig.IDM_limits_dicrete} shows the compilation of all the limits of the IDM benchmarks considered in this work. The available scenarios are represented by points in the relic abundance curves (red) that lie below the limits of LZ, H.E.S.S. (546 h of observation at the GC), and the projected CTAO (525 h of observation at the GC) limits. The indirect detection limits were obtained under the assumption of an Einasto profile. It is important to emphasize that these Galactic Center limits are inherently tied to the spatial distribution of the chosen dark matter density profile. For instance, assuming a cored density profile, such as a kiloparsec-sized Burkert distribution, can weaken the H.E.S.S. limits by up to two orders of magnitude \cite{HESS:2022ygk}. Conversely, due to the larger regions of interest mapped out by CTAO, the sensitivity deterioration in cored scenarios is much less severe, scaling down by a factor of only a few \cite{CTA:2020qlo}. Furthermore, while our baseline CTAO projections reflect an optimistic statistical-only setup, a full integration of realistic systematic errors and Galactic Diffuse Emission (GDE) contamination would conservatively degrade these projected baseline limits (by a factor of approximately 2 to 3 due to instrumental systematics at sub-TeV masses, and by up to a factor of 4 to 6 when factoring in astrophysical modeling uncertainties of the interstellar emission across the multi-TeV regime) as investigated extensively in Ref.~\cite{CTA:2020qlo}.

In the first place, the main IDM feature is evidenced as $\left< \sigma_{\text{ann}} \, v \right>$ lies in the $5 \times 10^{-26} - 10^{-25} \, \text{cm}^3 \text{s}^{-1} $ range, sometimes greater than the typical annihilation cross section. It is also possible to see the complementarity between direct detection and indirect detection limits: for small mass-splittings, $\Delta_+, \Delta_o \lesssim 1 \, \text{GeV}$, LZ exclusion dominates in the $0.5 \, \text{TeV} - 2 \, \text{TeV}$ mass range, while H.E.S.S. is stronger for $m_{DM}$ in $2 - 20\, \text{TeV}$. For greater mass splittings, H.E.S.S. always dominates over direct detection. H.E.S.S. excludes masses between $\approx 1 \, \text{TeV}$ and $ \approx 10 \, \text{TeV}$, for all benchmarks. This means that the most recent indirect detection data exclude most IDM scenarios from the Inner Galaxy Survey of the H.E.S.S. telescope~\cite{HESS:2022ygk}. Furthermore, even when considering the aforementioned systematic and astrophysical uncertainties, the projected CTAO limits are expected to exceed those of LZ and H.E.S.S. for virtually all masses, making it a definitive probe for the model.

\begin{figure*}[!ht]
    \centering
    \includegraphics[scale=0.7]{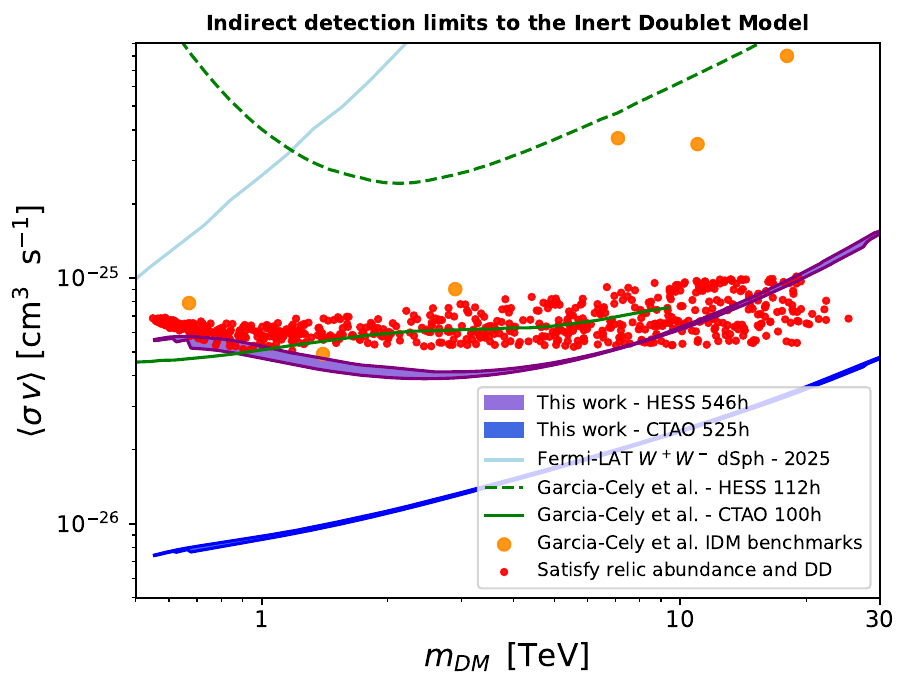}
    \caption{Sensitivity of gamma-ray observatories to the Inert Doublet Model. Points in red represent viable scenarios obtained in the random scan that satisfy unitarity, relic abundance, and direct detection limits. The blue and purple bands correspond to the CTAO and H.E.S.S. limits obtained from the maximum and minimum values of the collection of all obtained benchmarks, assuming an underlying Einasto profile. The CTAO projections reflect a statistical-only baseline scenario. Orange dots are the benchmark points, including the Sommerfeld enhancement computed in \cite{Garcia-Cely:2015khw}. The CTAO and H.E.S.S. limits used in their work are shown as solid and dashed green lines, respectively. We also included in light blue the Fermi-LAT limits for dwarfs. }
    \label{fig.IDM_limits_continuous}
\end{figure*}

The variability of branching ratios across benchmarks implies differences in indirect detection limits. To address that, the upper and lower limits of the benchmarks were treated as a \textit{band of exclusion}. This enables comparison of viable scenarios from the random scan with the indirect detection limits obtained from the discrete scan (see Appendix A for further discussion). Figure \ref{fig.IDM_limits_continuous} shows the comparison of the exclusion bands and the relic abundance points. The red points correspond to scenarios obtained from the random scan that satisfy relic abundance, unitarity, and direct-detection limits from LZ data. As in Figure \ref{fig.IDM_limits_dicrete}, thermal velocity-weighted annihilation cross sections are in the $5 \times 10^{-26} - 10^{-25} \, \text{cm}^3 \text{s}^{-1}$ range, which is greater than the typical WIMP value due to co-annihilation. Regarding indirect detection, all points above the bands are excluded; points below the bands remain viable; and points within the bands have annihilation cross sections near the observatory's sensitivity, so their exclusion will depend on their specific parameters. The conclusions are equivalent to the ones from the discrete scan: actual H.E.S.S. data (purple) using the Einasto profile excludes a great part of viable scenarios, including all DM masses $m_{DM}$ between $\approx 1 - 8 \, \text{TeV}$. Some scenarios in the $0.6 - 1 \text{TeV}$ mass range with $ \Delta_+ \lesssim 2 \, \text{GeV} $ and $ \Delta_o \gtrsim 5 \, \text{GeV} $ are within the band of H.E.S.S. exclusion. A significant number of points with masses in the $10 \, \text{TeV} \lesssim m_{DM} \lesssim 20 \, \text{TeV} $ range are under H.E.S.S. limits and are still valid scenarios of the Inert Doublet Model. These remaining high-mass scenarios generally have $\Delta_+ \gtrsim 8 \, \text{GeV}$, while $\Delta_o$ does not appear to be correlated in this region. The projected CTAO limits (blue) are almost one order of magnitude stronger than those from H.E.S.S., and they would exclude all viable scenarios. This means that CTAO will be able to probe all the DM parameter space of the Inert Doublet model with small mass-splitting ($\Delta_{o,+} = m_{A,H^\pm} - m_{H} \lesssim 10 \, \text{GeV} $). For completeness, we also included the Fermi-LAT limits for dwarf spheroidal galaxies for the $W^+W^-$ channel (light blue) \cite{Fermi-LAT:2025gei}, but it has no sensitivity to probe this high mass parameter space.  

Moreover, it is important to highlight the changes expected by the inclusion of the SE in the IDM. First, the relic abundance can deviate by up to 30 \% depending on whether the SE is taken into account or not \cite{Garcia-Cely:2015khw} \cite{Hisano:2006nn} \cite{Beneke:2014hja}. Second, for a broad region of the parameter space, the thermal cross section in the Galactic Center can be enhanced by factors as large as $10^4$ for masses in the range between 5 and 20~TeV~\cite{Garcia-Cely:2015khw}. The impact of this non-perturbative calculation exceeds the uncertainties associated with the early Universe. Consequently, a significant region of the parameter space with $m_H > 10 $ TeV is expected to be excluded by the HESS 546h data when the non-perturbative effects are included. On the other hand, scenarios with $m_H$ below a few TeV are not expected to change significantly. In Fig.~\ref{fig.IDM_limits_continuous}, we add to the plot the benchmark points (orange), including the SE as computed by \cite{Garcia-Cely:2015khw} and the CTAO (green) and HESS (dashed green) limits used there. As we can see, our limits are much stronger than the previously reported limits\footnote{Even when accounting for a re-scaling of the limits by the square root of the increase in observation time, our limits remain more than a factor of 2 stronger.} and show that the IDM model is already excluded for $m_H <8$ TeV by the current H.E.S.S. data, considering the Einasto profile. If those benchmark values of SE are taken into account, the model is strongly constrained in the whole high-mass regime except for possible fine-tunings. The remaining points will be explored by the CTAO, leading to a total probe of the model. 

\section{Conclusions}

In this work, we explored the phenomenology of the IDM, focusing on the ``high mass regime'' ($m_H \gtrsim 500 \, \text{GeV}$). We mentioned that, to achieve the correct relic density, we need a higher thermal annihilation cross section due to co-annihilation effects than in standard WIMP production. This leads to an exciting scenario in which the usual collider and direct-detection limits lose sensitivity, leaving room to explore the IDM's phenomenology through indirect detection. We showed that, after imposing all limits, including theoretical and direct and collider constraints, the most recent Inner Galaxy Survey by the H.E.S.S. telescope excludes nearly the entire viable parameter space of the IDM, choosing the Einasto halo profile. Furthermore, considering the most recent IRFs, we computed the projected limits for the CTAO and showed that it will probe virtually all remaining viable dark matter scenarios within the high mass regime of the IDM.


\section*{Acknowledgements}
We thank Enrico Bertuzzo, Edivaldo Moura, André Lessa, and Farinaldo Queiroz for discussions and comments.
Authors are supported by the S\~{a}o Paulo Research Foundation (FAPESP) through grants number 2019/14893-3, 2020/00320-9, 2021/01089-1. AV is supported by FAPESP grant No 2024/15560-6. LRJ is supported by Capes through grant number 88887.684414/2022-00 and by FAPESP through grant number 2022/01962-0. CNPq supports AV and CS through grant numbers 406718/2025-3, 314955/2021-6 and 304944/2025-4, respectively. The authors acknowledge the National Laboratory for Scientific Computing (LNCC/MCTI, Brazil) for providing HPC resources for the SDumont supercomputer (http://sdumont.lncc.br). This work was conducted in the context of the CTAO Dark Matter Working Group.

\appendix

\section{Branching ratios of annihilation in the Inert Doublet Model}

\begin{figure*}[!ht]
    \includegraphics[width=\textwidth]{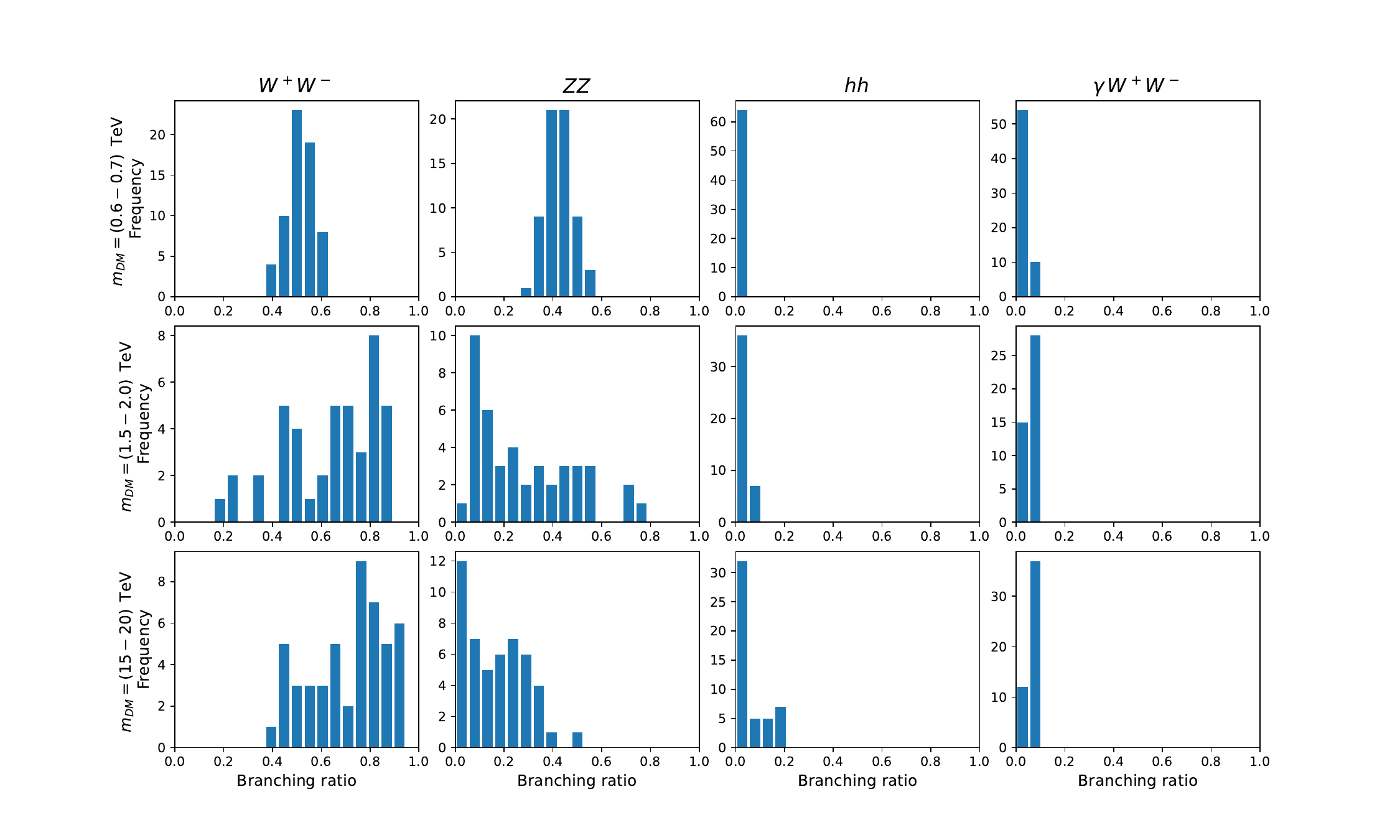}
    \caption{Distribution of branching ratios $B_i$ for the main channels of DM annihilation in the IDM: $W^+ W^-$, $Z Z$, $h h$, and $\gamma W^+ W^-$. Each column represents one of these channels, and each row is a sample of scenarios considered.}
    \label{fig.IDM_hist}
\end{figure*}

In the IDM, indirect detection limits change according to the scalar mass $m_{DM}$ and the mass splittings $\Delta_{+/o}$. This slight difference appears due to the difference in the branching ratios $B_i$. This means that understanding how branching ratios depend on the IDM scenarios is important for interpreting the indirect detection results. The main channels in the high mass regime were found to be the bosonic channels $W^+ W^-$, $Z Z$, and $h h$, as pointed out by previous studies \cite{Queiroz:2015utg}. The quark channel $tt$ is the most relevant among the fermion states, but its contribution is at most $\approx 1 \%$ in our scans. Another channel that can play a role is the EW radiative emission $H H \rightarrow \gamma W^+ W^-$ (see Refs.~\cite{Garcia-Cely:2013zga} and \cite{GarciaCely:2014jha} for a detailed discussion about this channel). The distributions of branching ratios for these four channels are shown in Figure \ref{fig.IDM_hist} for three ranges of mass sampled from the random scan: $ 0.6 \, \text{TeV} < m_{DM} < 0.7 \, \text{TeV} $, $ 1.5 \, \text{TeV} < m_{DM} < 2.0 \, \text{TeV} $ and $ 15 \, \text{TeV} < m_{DM} < 20 \, \text{TeV} $. These histograms show that $W^+ W^-$ and $Z Z$ states dominate the annihilation channels, accounting for almost 90\%. Both channels have branching fractions around 50\% in the first sample (first row of Figure~\ref{fig.IDM_hist}), and their distributions broaden at higher masses. The $W^+ W^-$ state is dominant over $ZZ$ for $m_{DM} > 1.5 \, \text{TeV}$ for most scenarios. Regarding other channels, the Higgs $h h$ one does not surpass $ \approx 20\%$ of contribution, while the EW radiative $\gamma W^+ W^-$ correction reaches contributions of at most $\approx 10 \%$. That variability in branching contribution also applies to the discrete scan, which explains the differences in indirect detection limits.

\section{Derivation of H.E.S.S. and CTAO limits}
\label{app:lim}

For the limits obtained from H.E.S.S., ON and OFF counts from real data corresponding to 254 hours of observation were utilized \cite{abdallah2016search}. Under the assumption that the background rate remains stable, an extrapolation of these data to 546 hours was performed to match the cumulative observation time of the 2014-2020 Inner Galaxy Survey \cite{HESS:2022ygk}. The IRFs were extracted from Ref. \cite{lefranc:tel-01374541}. The official H.E.S.S. work \cite{HESS:2022ygk} reported no significant signal indicating dark matter annihilation in the GC. Utilizing the profile likelihood method, they excluded annihilation cross sections near the thermal relic value under the assumption of an Einasto or NFW halo model. For CTAO, the dataset was simulated using the updated IRFs available on the CTAO consortium website. We employ the more recent Alpha Configuration IRFs, downloaded via the \textit{prod5 version v0.1} repository. A total of 525 hours of observation time is considered, and our baseline results are validated against the official results provided in the 2021 Consortium publication \cite{acharyya2021sensitivity}.  
\begin{figure}
    \centering
    \includegraphics[width=1\linewidth]{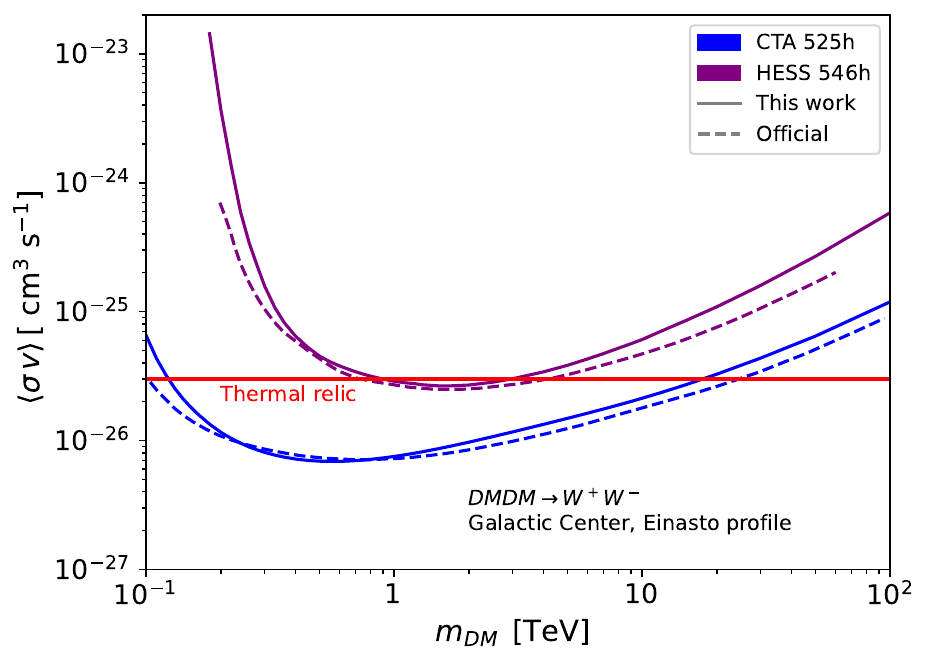}
    \caption{Expected exclusion limits on the dark matter annihilation cross section in the $\left< \sigma_{\text{ann}} \, v \right> \times m_{DM}$ space for the bosonic $W^+ W^-$ channel. The limits are shown for the HESS 546h of data (purple lines) and the projected performance of 525h of CTAO (blue lines), both with the GC as the target. The results of this work (solid) and the official ones (dashed) are plotted together for comparison. The thermal relic cross section $ \left< \sigma_{\text{ann}} \, v \, \right> = 3 \times 10^{-26} \, \text{cm}^3 \text{s}^{-1} $ is shown in red \cite{profumo2019introduction}. }
    \label{fig:comp}
\end{figure}

The upper limits on the annihilation cross section $\left\langle \sigma_{\text{ann}} \, v \right\rangle$ as a function of the dark matter particle mass $m_{DM}$ for the primary $W^+ W^-$ channel are shown in Figure \ref{fig:comp}. In the plot, the projected CTAO limits are shown in blue, while the H.E.S.S. limits are shown in purple. Only statistical expected limits were taken into account (no systematic errors). We adopt a benchmark cuspy Einasto profile \cite{Acharyya2021CTA} and used the PPPC4DMID package to build the spectra \cite{Cirelli2011PPPC,Cirelli2012PPPC-Erratum}.  In all cases, the baseline sensitivities derived in this work (solid lines) demonstrate an excellent agreement with the official limits published by both collaborations (dashed lines).








\end{document}